%% file: paper.tex
\documentclass[conference,compsoc]{IEEEtran}
\usepackage[usenames,dvipsnames,svgnames,x11names]{xcolor}
\usepackage[lofdepth,lotdepth]{subfig}

\newcommand{\ysnoted}[1]{} 

\usepackage[normalem]{ulem} 

\usepackage[nocompress]{cite}
\usepackage{lipsum}
\let\OLDthebibliography\thebibliography
\renewcommand\thebibliography[1]{
	\OLDthebibliography{#1}
	\setlength{\parskip}{0pt}
	\setlength{\itemsep}{0pt plus 0.3ex}
}

\usepackage{filecontents, pgffor}
\input{solidity-highlighting.tex}

\usepackage{listings}

\lstdefinelanguage{XML}
{
	basicstyle=\ttfamily\footnotesize,
	morestring=[b]",
	moredelim=[s][\bfseries\color{Maroon}]{<}{\ },
	moredelim=[s][\bfseries\color{Maroon}]{</}{>},
	moredelim=[l][\bfseries\color{Maroon}]{/>},
	moredelim=[l][\bfseries\color{Maroon}]{>},
	morecomment=[s]{<?}{?>},
	morecomment=[s]{<!--}{-->},
	commentstyle=\color{gray},
	stringstyle=\color{blue},
	identifierstyle=\color{red}
}
\usepackage{moreverb}

\usepackage[nounderscore]{syntax}

\usepackage[pdftex]{graphicx}
\graphicspath{{./figures/}}
\DeclareGraphicsExtensions{.pdf}
\usepackage{subfig} 
\newcommand{\tuple}[1]{\mbox{$\langle #1 \rangle$}}

\usepackage[cmex10]{amsmath}
\usepackage{amssymb}
\usepackage{mathtools}
\usepackage{amsthm}
\usepackage{amsfonts}

\usepackage{algorithmicx}
\usepackage{algpseudocode}
\usepackage[ruled]{algorithm}
\definecolor{light-gray}{gray}{0.75}
\algrenewcommand{\algorithmiccomment}[1]{\hskip3em{{\footnotesize \textcolor{light-gray}{$\blacktriangleright$}}} #1}
\usepackage{multirow} 
\usepackage{rotating} 
\usepackage{booktabs} 
\usepackage{colortbl} 
\usepackage{tablefootnote} 
\usepackage{array}
\newcolumntype{R}[1]{>{\raggedleft\let\newline\\\arraybackslash\hspace{0pt}}m{#1}}

\usepackage[pdftex,colorlinks=true,urlcolor=blue,citecolor=blue]{hyperref}

\usepackage{xspace}

\let\labelindent\relax
\usepackage{enumitem}

\usepackage[english]{babel}
\usepackage{blindtext}

\usepackage{balance}

\makeatletter
\newcommand*{\balancecolsandclearpage}{%
  \close@column@grid
  \cleardoublepage
  \twocolumngrid
}
\makeatother

\begin{document}
%
\title{Smart Contracts for Multiagent Plan Execution in Untrusted Cyber-physical Systems}

\author{\IEEEauthorblockN{Anshu Shukla, Swarup Kumar Mohalik, Ramamurthy Badrinath}
\IEEEauthorblockA{Ericsson Research, Bangalore, India\\
Email: \{anshu.shukla, swarup.kumar.mohalik, ramamurthy.badrinath\}@ericsson.com}
}

\maketitle

\begin{abstract}
Intelligent Cyber-physical systems can be modeled as multi-agent systems
with planning capability to impart adaptivity for changing contexts.
In such multi-agent systems, the protocol for plan execution must result in 
the proper completion and ordering of actions in spite of their distributed execution. 
However, in untrusted scenarios, there is a possibility of agents not respecting
the protocol either due to faults or due to malicious reasons thereby resulting in plan failure. 
In order to prevent such situations, we propose to implement the execution of agents through
smart contracts. This points to a generic architecture seamlessly integrating intelligent 
planning-based CPS and smart-contracts.
\end{abstract}

\IEEEpeerreviewmaketitle

\section{Introduction}
Cyber-physical systems (CPS)  comprise sensing and actuation capabilities
controlled typically through software agents. A view of intelligence in these systems attempts to equip the systems with self-adaptivity so that the systems
can adapt to dynamic changes in the environment and system capabilities automatically.
Such self-adaptivity can be imparted by integrating artificial intelligence (AI) planning subsystems in the software agents. Frameworks along this philosophy have been studied in the autonomous systems literature ~\cite{mape-k}

Extending the above approach to distributed CPS, one naturally has multi-agent planning based systems where both plan synthesis and execution need coordination between the agents. The plans are maintained in data structures that can be global and is orchestrated by a central agent, or in local data structures and are executed through agreed protocols between the agents. These are the central themes of multi-agent planning literature \cite{deWeerdt:2009}.
In this paper, we assume that the plans have already been synthesized (global or local) and concern ourselves only with the execution phase.

A hidden assumption in multi-agent planning is that once the plan is synthesized
and the protocol has been agreed upon, the agents execute strictly according to
the protocol. However, in an untrusted environment, this assumption may not be 
tenable. Either due to malicious reasons or selfishness, or due to faults, the agents may not be able to adhere to the protocols, which may lead to one or more number of anomalies such as out-of-order execution, untimely initiation and improper completion of actions etc. In critical CPS, the cost of resulting failures being very high, it is 
necessary to devise a mechanism where agents cannot deviate from the established and agreed-upon protocol.

Towards this, we propose a multi-agent plan execution framework based upon Smart contracts on blockchains. The design of smart contracts for the plans along with the properties transparency, authenticity and immutability enforced by blockchains ensure
the desired execution of plans in spite of the untrusted environment. The structure of the smart contracts is dependent upon the multi-agent plan execution architecture (centralized or decentralized execution). 

To the best of our knowledge, this is the most direct application of smart contracts and blockchains in AI planning. So far, blockchains have been used mainly to store data, AI models and algorithms to ensure integrity in untrusted environments. Controlling the plan execution via validated smart contract transactions seems to be a completely novel contribution.

The paper is organized as follows. In the following section, we give some basic introduction to planning, multi-agent planning, blockchains and smart contracts. Section 3 describes the system model formally and outlines the issues of plan execution in untrusted environments. We give the solution based on smart contracts in Section 4. The review of literature in the intersection of smart contracts AI is given in Section 5. Section 6 concludes the paper with a summary and number of possible extensions.

\section{Background}
Cyber-physical systems(CPS) consist of a number of physical {\em devices} that can be controlled by one or more {\em agents}. Typically, the devices expose a set of actions through an API that is utilized by the agents. In autonomous CPS, the agents are intelligent i.e. they adapt to changes in the environment and/or requirements and control the devices differently to achieve changed goals. In AI literature, planning - plan synthesis and execution - is one of the approaches of endowing agents with intelligence. 

\subsection{Planning}
AI Planning is an umbrella of techniques which derives a sequence of actions (called a Plan) that can lead from an 
initial state to a specified goal state. Each {\em state} is described by a set of predicates about the objects of the 
world and their relationships. {\em Actions} describe the change they cause to a state of world. They are specified through 
\tuple{precondition, effect} pairs that are boolean combination of predicates. An action is enabled 
in a state only when its precondition is satisfied in that state. The execution of the action leads to a new state where 
its effect holds.

A {\em planning domain} specifies the predicates (and functions) necessary to capture the states
of the domain under consideration. It also specifies the actions through their
precondition and effect pairs. On the other hand, a {\em planning problem} captures the planning requirements 
through  a specific state of the domain as initial state and a specific goal state.  For more details on planning see \cite{russell-norvig}.

\begin{figure}[t]
	\centering
	\includegraphics[width=0.4\columnwidth]{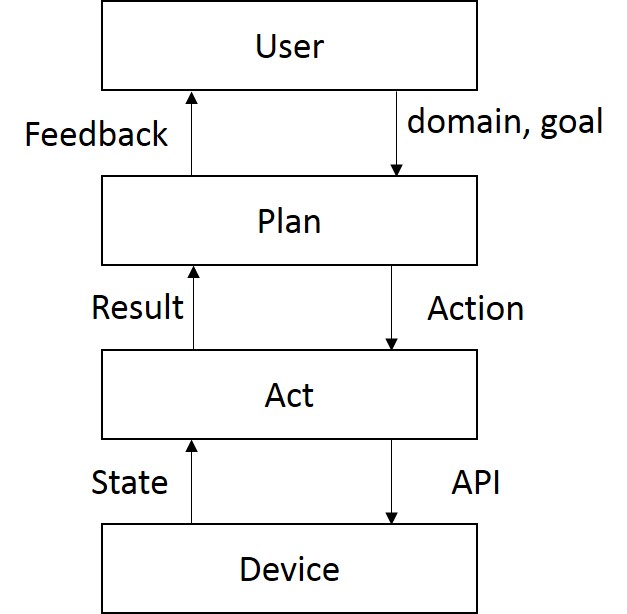}
	\caption{Architecture of planning based intelligent agent}
	\label{fig:int-agent}
\end{figure}

AI planning involves plan synthesis and plan execution. The typical architecture of an intelligent agent is given by the basic sense-plan-act loop or its variations, such as MAPE~-~K~\cite{mape-k}. Fig.~\ref{fig:int-agent} gives the schema of a bare-boned planning based intelligent agent. Note that the {\em plan} component may just acquire a plan, say from a user or plan database, instead of having to actually compute it.

Planning algorithms derive {\em plans}, which are a partially ordered set of nodes labeled with actions. A plan is said to be {\em properly} executed when for each action $a$ in the partial order, (1) $a$ is executed only when the actions on which it is dependent - denoted {\em dep(a)} - have already completed execution, (2) the pre-conditions of $a$ are satisfied and (3) when $a$ completes execution, the effects of $a$ hold true. Otherwise one could conclude that either the action did not execute properly or the modeling of the effects of the action was wrong. Essentially, the actions in the partial order plan should execute in a topological order with additional constraints imposed by preconditions and effects.

\subsection{Multi-agent Planning}
Multi-agent planning is a natural extension of the basic planning problem where there are more than one agent. The extension gives rise to a number of variations such as: whether the agents are cooperative or competitive, whether planning is centralized or not,  whether communication is reliable or not\cite{deWeerdt:2009} etc.

The basic problem in multi-agent agent planning is the distribution of plans over the agents. These plans must be executed in a coordinated manner so as to achieve the system goals. Therefore, the agents must follow certain protocols that are part of the plan generation. 

In the present paper, we address two execution architectures for multi-agent planning. 
The first one is a centralized architecture where a {\em scheduler} agent has the full plan and orchestrates the execution by dispatching them to the agents in a timely manner(Fig.~\ref{fig:cent-plan-execution}). The second one is where each agent acquires its plan and the execution is done in a coordinated manner through communication between the agents(Fig.~\ref{fig:dist-plan-execution}).

\subsection{Blockchain \& Smart Contract}
A \emph{blockchain} is a distributed, decentralized ledger that maintains a common set of updated and secure records of  transactions across many computers based on peer-to-peer networks and cryptographic algorithms. Each peer in the network validates the transaction against the rules set in the specific blockchain network~\cite{stellar:white:2017,iota:white:2017,cachin:hyperledger:2016}. A certain number of valid transactions are grouped to form a \emph{block}. Each block has a reference to the preceding block in the blockchain allowing the participants to confirm the integrity of the blocks all the way back to the original genesis block.
A copy of blockchain is stored on every machine participating in the blockchain network and is synchronized among the peers continuously. Thus, it is a permanent ledger that all peers on the network use for verification, coordinating actions and to reach an agreement in an auditable manner.

A smart contract~\cite{buterin:2014} is essentially a collection of code and data representing some business logic running on a blockchain. The smart contract resides at specific address on the blockchain. Solidity~\cite{solidity} is one of the commonly used languages for coding smart contracts.
The basic example for smart contract can be just a data update operation on blockchain; e.g., updating the account balance after validating that account has enough money before debiting it. As a more complex example, in the smart logistics domain one may dynamically adjust the shipping cost based on the time of the request. The ledger will have agreed terms by both parties coded as smart contracts. The appropriate funds incorporating the dynamic delivery charges are transferred automatically based on delivery time of item.

\section{System Model}
In this section, we describe the formal model of the multi-agent planning based CPS and define the problems to be addressed. 

The CPS under consideration consists of a set of devices ${\cal D}$ where each device $d$ is associated with a number of actions $Act_d$ exposed through a REST-based API. We denote the set of all the actions from the devices as $Act$.
The REST-based API abstracts the underlying communication and device libraries and is not relevant to the scope of the present paper. We denote by $execute(a)$ a call to the API that executes an action $a$ on the corresponding device.

Each device $d$ has a number of sensors collecting data about the state of the device. We model the state of the devices via a set of predicates whose values can be decided by processing the sensor values. The set of all predicates is denoted as $\cal P$ and the predicates associated with a device $d$ is denoted as ${\cal P}_d$. Similar to the actions, the values of the predicates are assumed to be accessible through a REST-based API. We denote by $getVal(p)$ the REST-based call to get the value of the predicate $p$.

One or more devices have an associated $agent$ which is implemented in software. The set of all agents is denoted as $AG$. The spatial distribution function $loc : Act~\rightarrow~AG$ maps each action of a device to the associated agent. At the basic level, each $agent$ can call the $getVal(\cdot)$ API to access the value of a predicate and call $execute(\cdot)$ to execute an action in a device. Moreover, each $agent$ has the mechanisms to communicate with each other or with a separate central agent to exchange information. For the purpose of the paper, the information to be exchanged is essentially about the completion of an action by the device.

To assist automatic synthesis and control plan execution, the actions are annotated with two sets of predicates: {\em pre-condition}, which should be true for the action to be enabled and {\em effect}, which must hold as a result of action execution. Given an action {\em a}, we denote the pre-condition and effect of the action as {\em precond(a)} and {\em effect(a)} respectively.

An initial state $init$ is specified as a conjunction of state predicates from $\cal P$ and a $goal$ state is specified as a boolean expression over $\cal P$ as well. A plan for the problem $(init,~goal)$ is derived as a partial order over $Act$. 

A proper execution of the plan starts at the $init$ state, executes the actions in the plan in topological order and finally results in a state satisfying $goal$. The agents check the precondition and effect constraints by accessing the predicate values through $getVal(\cdot)$. The actual execution of the actions is done through the $execute(\cdot)$ calls. In addition, for each action $a$ to be executed by the agent $loc(a)$, the agent needs to know that the actions in $dep(a)$ have already completed execution. Since these actions in $dep(a)$ might be with other agents, the action completion events must be notified through the communication mechanism between agents. This leads to two natural architectures for distributed plan execution.

\begin{figure}[ht]
	\centering
	\includegraphics[width=0.99\columnwidth]{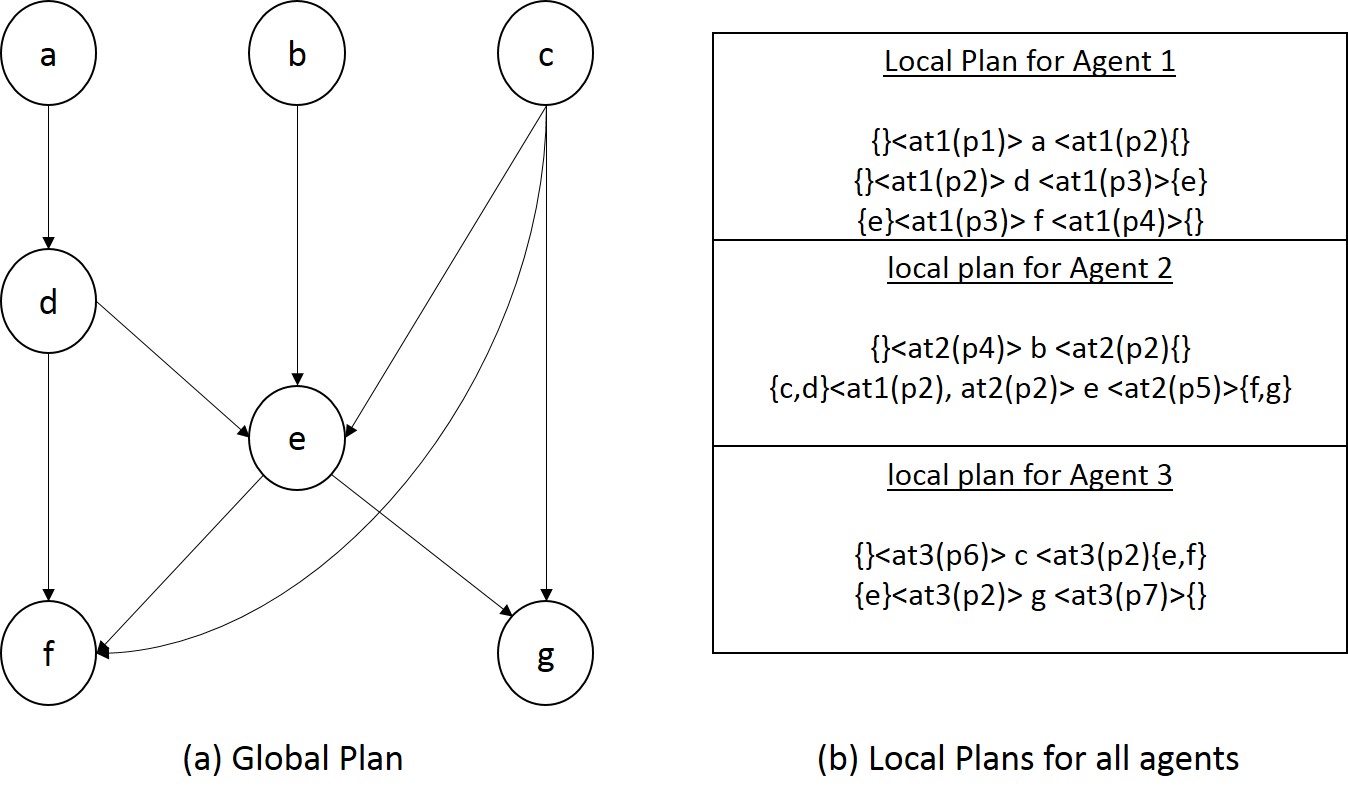}
	\caption{Global and local plans}
	\label{fig:global-local-plan}
\end{figure}

\subsection{Distributed Plan Execution in Centralized Mode}
In the centralized mode of plan execution, there is an external $Scheduler$ agent who schedules the execution of actions taking into account $dep(\cdot)$ of actions. The device agents are only responsible for the verification of precondition and effects and dispatch of actions through the action APIs. The general schema is depicted in Fig.~\ref{fig:cent-plan-execution}. 

The Scheduler implements the topological order of the plan by maintaining a datastructure $completedActionsList$ to record the completion of execution of actions as notified by the agents. When the dependencies of an action $a$ is detected to be fulfilled by inspecting the $completedActionsList$, Scheduler sends a request for dispatch to the agent $loc(a)$. The device agent accesses the state of the devices through the call $getVal(precond(a))$ and if it is true, issues the $execute(a)$ call. When the action is completed successfully (result of the $execute()$ call), the agent sends the completion message to the Scheduler which updates the $completedActionsList$. The interactions continue till all actions in the plan are completed ($completedActionList~==~actions~in~the~plan$).

\subsection{Distributed Plan Execution in Decentralized Mode}
\begin{figure}[t]
\centering
\subfloat[][Centralized mode]{
	
	\includegraphics[width=0.70\columnwidth]{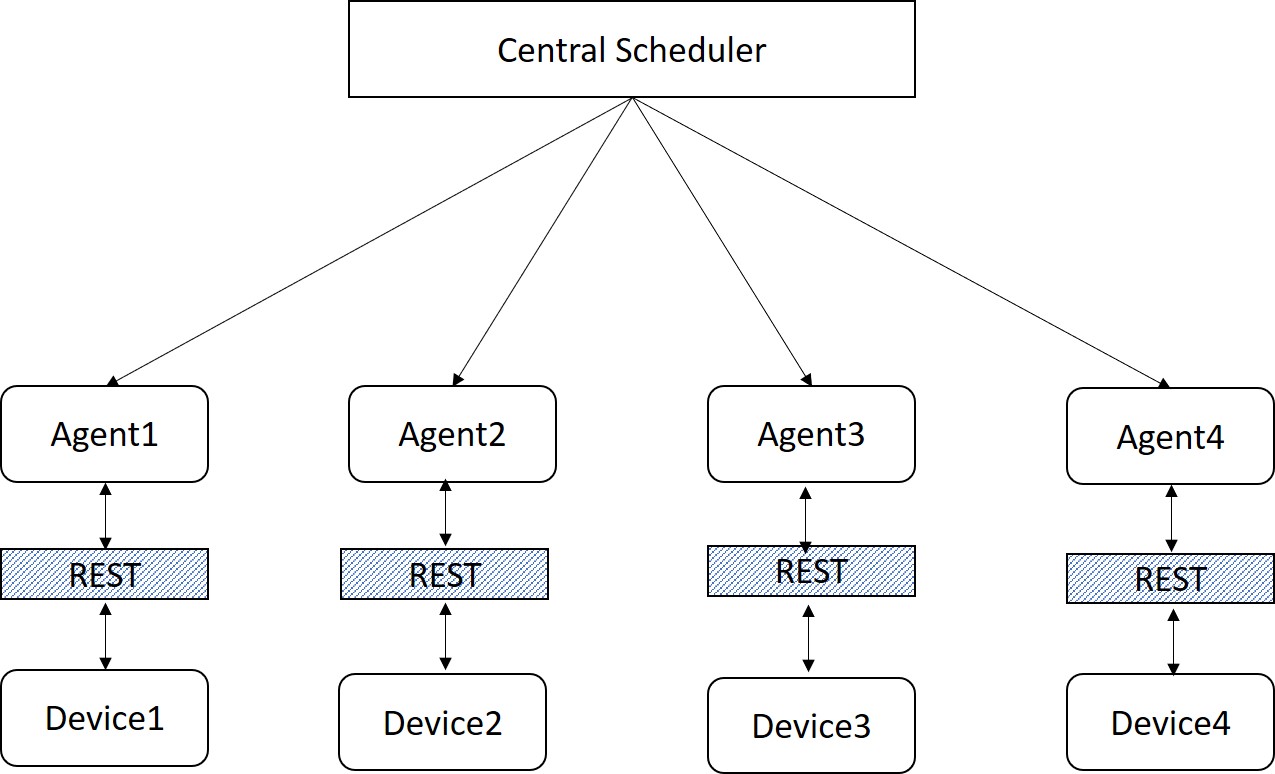}
	\label{fig:cent-plan-execution}
}
\qquad
\subfloat[][Decentralized mode]{
	\includegraphics[width=0.70\columnwidth]{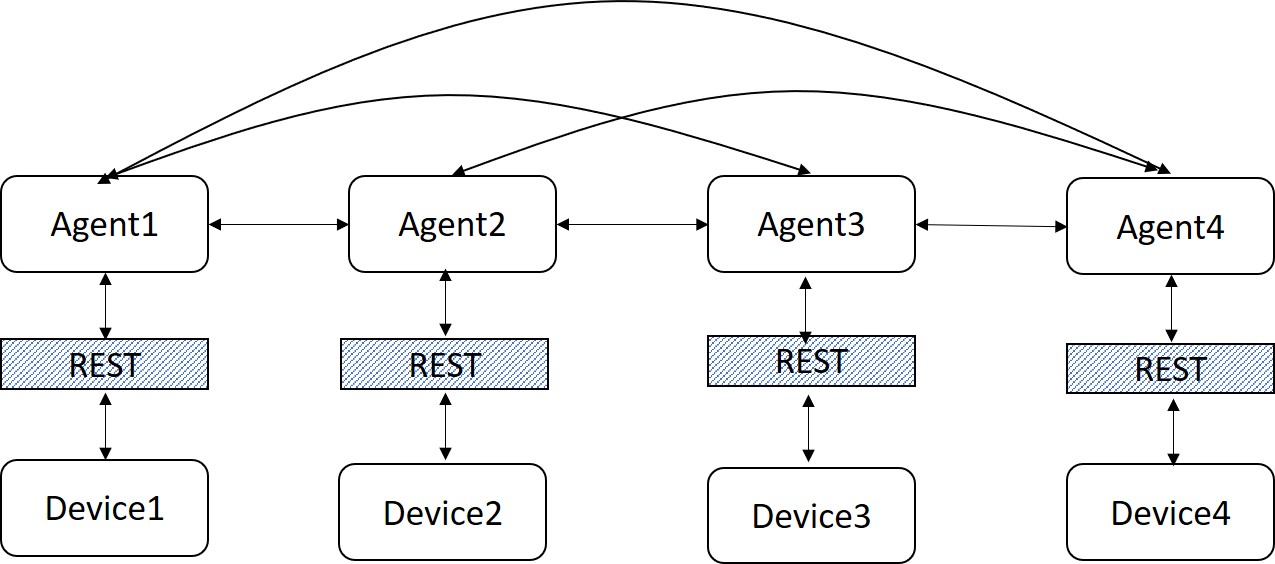}
	\label{fig:dist-plan-execution}
}
\caption{Architectures for distributed plan execution. \label{fig:arch-dpe}}
\end{figure}

In the decentralized mode for distributed plan execution, there is no single scheduler. The agents communicate and coordinate with each other to execute a plan (Refer Fig.~\ref{fig:dist-plan-execution}). 
Each agent maintains a $completedActionsList$ locally. It also maintains a local plan which is a set of local actions $\tuple{In(a), precond(a), a, effect(a), Out(a)}$, where $a$ is the label of the action, $In(a)$ is the set of actions on which $a$ is dependent and $Out(a)$ is the set of actions that are dependent upon $a$. Given a global partial order plan, it is easy to derive the local plans for each agent. 

Fig.~\ref{fig:global-local-plan} shows an example with three agents where Agent1 operates the actions $\{a, d, f\}$, Agent2 operates $\{b, e\}$ and Agent3 operates $\{c, g\}$. The preconditions and effects of the actions are specified through a bunch of predicates named $atX()$. Fig.~\ref{fig:global-local-plan}a shows the global partial order plan.
Corresponding local plans for the agents are shown in
Fig.~\ref{fig:global-local-plan}b. For example, for the action labeled $e$, the in-dependency is $IN(e) = \{c, d\}$ and the out-dependency is $Out(e) = \{f ,g\}$.

The local plan execution proceeds as follows: when an action $a$ is completed, the information is communicated to the agents in $Out(a)$ i.e. the agents whose actions have $a$ in their dependency. Then, each agent updates its $completedActionsList$ and decides whether it can schedule some action for execution by checking against $In(a)$. Only after that, the agent checks the precondition through $getVal(precond(a))$, calls the $execute(a)$ function and then again checks the effect through $getVal(effect(a))$ in that order.

\subsection{Issues with multi-agent plan execution in an untrusted environment}
The distributed plan execution mechanisms described above (both centralized and de-centralized) implicitly assume that the agents are trusted and collaborative. Therefore, plan execution can be implemented using
simple datastructures and communication mechanisms. However, in highly distributed CPS's with mobile entities entering and exiting the system dynamically, such assumptions may be misplaced. In the following, we discuss the points of failure in distributed plan execution in untrusted environments where agents may be faulty or malicious.

\begin{enumerate}	
\item An agent may not execute the action it is supposed to execute. This will result in blocking the dependent actions in the same or other agents, or not resulting in the desired effects, consequently leading to plan failure.
\item An agent may execute a local action out of order even though the dependencies are not yet fulfilled. This can happen when such early execution is beneficial to the agent but leads to failure in achieving the system goal (a la "tragedy of commons").
\item An untrusted agent may call the $execute()$ API  even though the preconditions do not hold. This may be because of malicious or greedy reasons on part of the agent since the agent might have predicted that such preemptive execution might lead to system failure (e.g. baking a cake without pre-heating the oven).
\item Agents may notify the completion of actions even though they might not have issued the API corresponding to the action or even though the effects were not satisfied. This may be because an agent may profit by "shirking" its responsibility.
\end{enumerate}

In the rest of the paper, we describe how one can implement the agents in a distributed plan execution system with smart contracts such that the above-mentioned issues are prevented.

\section{Distributed Plan Execution with Smart Contracts}
The main idea is to coordinate the steps in distributed plan execution through smart contracts. Then, the semantics of smart contracts and the underlying blockchain ensures that the (in)actions of the agents are securely recorded so that agents violating the smart contract (hence the intended plan execution) can be determined unequivocally. The associated penalty of contract violation then forces the agents
to follow the contract resulting in proper plan execution.

\subsection{Smart contracts for plan execution in centralized mode}
\begin{figure}
	\centering
	\includegraphics[width=60mm]{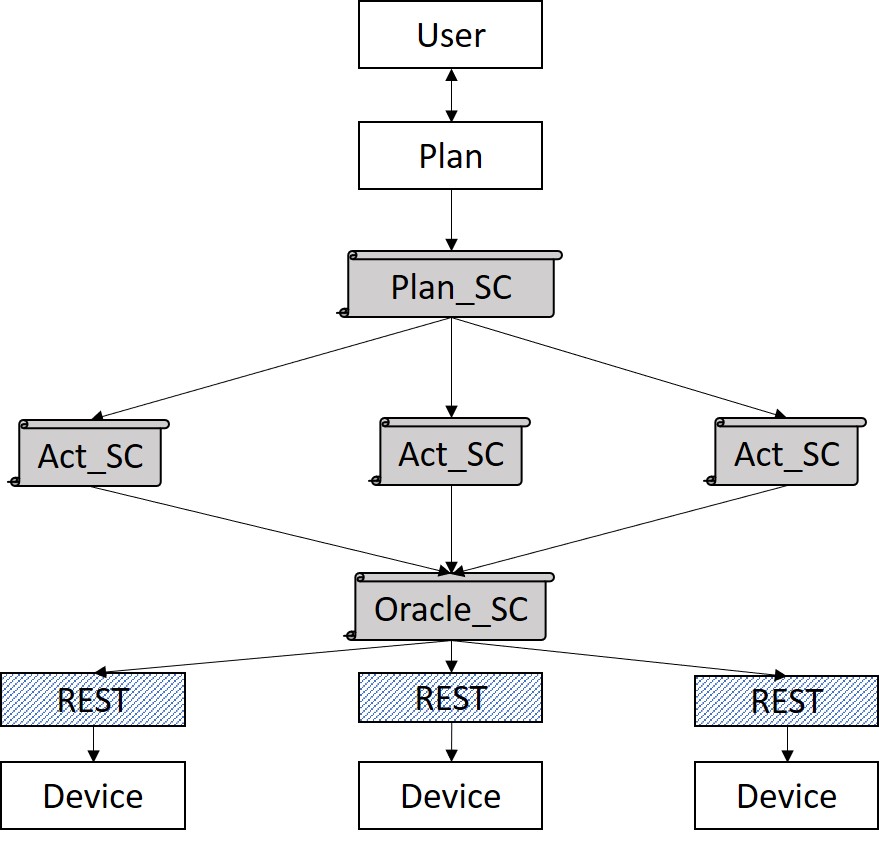}
	\caption{Smart contract based Distributed plan execution in centralized mode. \label{fig:cent-plan-exec-sc}}
\end{figure}

The architecture of the Smart Contract enhancement is shown in Fig.~\ref{fig:cent-plan-exec-sc}.  There is a smart contract ($Plan\_SC$) which is the Scheduler agent that orchestrates a given plan, a bunch of smart contracts ($Act\_SC$) each responsible for getting the predicate values and getting the actions executed. Since these actions need to access the datasources external to the underlying blockchain, the operations are executed as transactions on an Oracle smart contract($Oracle\_SC$)(explained in the following). In addition, for ease of information access, we use a Register smart contract($Register\_SC$) that essentially stores a mapping from the actions to the corresponding Act SC. 

We give the schema of the smart contracts in Listing ~\ref{lst:centralsc}. Complete code is given in the Appendix\footnote{The code is attached only for ready reference. To respect page limits, we will just provide a link to the code in a publicly available archive in the final version.}.

\begin{lstlisting}[language=Solidity, caption={\emph{Smart Contracts for Plan Execution in Centralized Mode}},label={lst:centralsc},captionpos=b,numbers=none,escapechar=\$]

contract Register_SC {
    event logDeployedContract(address tenant,string s);
 	mapping(uint256 => Act_SC) action_scAddrMap;
    function setAct(uint[] memory action_names) public { $\dots$ }
    function getAct(uint256 action_name) public returns(Act_SC act_scAddr) {$\dots$}
    }

contract Plan_SC{
    event loglist(uint256[] x,string s);    
    event logexist(uint256[] x,uint s,bool b);
    
    function Plan_SC(){//DAG Creation
    :
    }
     function dispatchNext(Register_SC register_addr, uint256 Action_id) returns(bool _success){ $\dots$ }

import "github.com/oraclize/ethereum-api/oraclizeAPI.sol";
contract Act_SC  is usingOraclize{
    	struct Action{
        uint256  Action_id;
        string Action_URI;        
        bytes32 [] preCond,effects;       
    }    
    Action private action_1,....,action_6;
    mapping(uint256 => Action) actionid_ActionMap;

    function Act_SC(uint[] memory action_names){$\cdots$}
    function getPreCondition(uint256 Action_id) public returns (bytes32[] preCond){$\cdots$}
    function getEffect(uint256 Action_id) public returns(bytes32[] effect) {$\cdots$}
    function checkPreCondition(bytes32[] preCond,bytes32[] preCond_orac) returns(bool cond){$\cdots$}
    function checkEffect(bytes32[] eff,bytes32[] eff_orac) public returns(bool cond){$\cdots$}
    function execute(uint256 Action_id) external returns(bool success){$\cdots$}
}   
\end{lstlisting}
\subsubsection{Outline of the smart contracts}
The Plan smart contract acquires a plan as a part of its constructor. In the listing, we have hardcoded the plan, but it can be obtained from the user or application or from some service. The plan is represented as a list of lists, the head of each sub-list denoting an action $a$, and the rest denoting $In(a)$. Whenever there is a completion notification for an action, this is deleted from all the lists. Thus, when the sublist is a singleton, it is known that its in-dependencies have been already completed and the head of the list is taken up for execution by calling $dispatchNext$.

There are multiple instances of the Act smart contract each corresponding to an agent. For each action that the Plan smart contract wants to dispatch, it must know the corresponding Act smart contract address. This information is maintained in a mapping (actions $\rightarrow$ Act SC instance address) in the Register smart contract. The $setAct$ function is used to create the mapping and $getAct$ function is used to extract the address of the Act smart contract instance.

Taking advantage of the shared blockchain, the different instances of Act smart contract instance share a common $completedList$ of actions. When the action is executed by an Act instance and the $effect$ has been verified, it updates the list. The Plan smart contract accesses the $completedList$ to see if the execution was successful and whether it can update the plan datastructure (list of lists).

Note the import statement in the Act smart contract: {\em import "github.com/oraclize/ethereum-api/oraclizeAPI.sol";} and the definition of Act smart contract as derived from Oraclize: {\em contract Act\_SC is usingOraclize}. This prepares Act SC to interface with Oraclize with the $\_\_callback(bytes32~myid,~bytes32[]~result)$ function. In the main function $execute()$, the Act smart contract checks the precondition after getting the predicate values from the oracle, calls the device execution through the Oracle (this is skipped in the code since we are just faking the execution, but in real application too, this is no different from the Oracle calls), and checks the effect after which it updates the $completedList$. Any failure in any of the above steps results in aborting the transaction, which can ultimately be cascaded to the user or application.

\subsubsection{Oracle SC from Oraclize}
The Oraclize~\cite{oraclize} solution guarantees that the data fetched from the original data-source is genuine and untampered. This is achieved by creating a document called authenticity proof for the data returned from the sensors. These proofs can be build using different technologies like Trusted Execution Environments e.g. Intel SGx\cite{costan:intelsgx:2016} and auditable virtual machines. The communication between smart contract and Oraclize is asynchronous in nature and happens as follows: Every transaction broadcast by an agent participating in the blockchain will have special instruction which manifest to Oraclize. The 3rd party service running the oraclize will be constantly monitoring the blockchain for such requests of sensor Data. As per the request, Oraclize will collect the result, sign and execute the \texttt{\_callback} function in smart contract. It also follows an "If This Then That" logical model. For example, it could continuously check for a condition and notify when the condition has been met. The oracles can be both Inbound, to get data inside the blockchain or Outbound, to notify an actuator outside of the blockchain about an event.

\subsubsection{Deployment and Running}
First the oracle smart contract is deployed. Using its address, the Act SC instances are deployed. User then deploys the Register SC that, in turn, calls the Act SC functions to populate the mapping (actions $\rightarrow$ Act SC instance address). Lastly, the Plan SC is deployed. After deployment, user or application can upload a plan to the Plan SC through the constructor. We plan to provide a {\tt plan.setPlan(plan)} function to upload plans during operation as well which can help in replanning. The {\tt plan.startExec()} starts off the execution of the plan by repeated calls to the {\tt dispatchNext()} function.

\begin{figure}[ht]
	\centering
	\includegraphics[width=0.99\columnwidth]{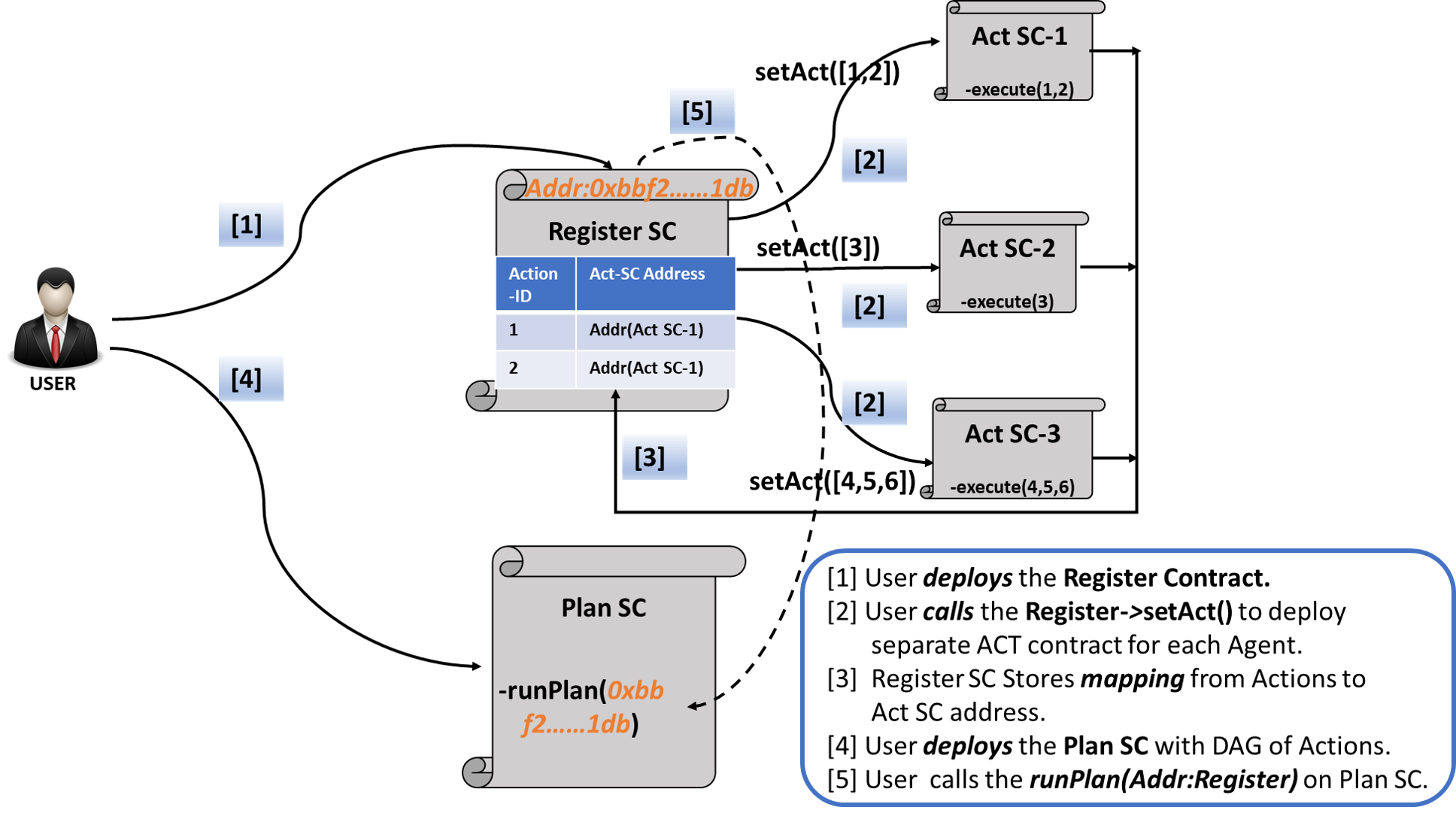}
	\caption{Deployment and Execution of smart contracts for plan execution}
	\label{fig:deploy-exec-sc}
\end{figure}

The function {\tt plan.dispatchNext()} selects actions whose dependencies have been fulfilled. These actions are then passed to {\tt Act.execute()}. Preconditions of these actions are checked by obtaining the predicates from the Oracle SC. After the action is invoked through the corresponding API, effect predicates are checked. Any exception in any of the steps leads to call of {\tt plan.abort()}.
These steps are illustrated in a pictorial manner in Fig.~\ref{fig:deploy-exec-sc}. 

\subsection{Smart Contracts based plan execution in decentralized mode}

\begin{figure}
	\centering
	\includegraphics[width=80mm]{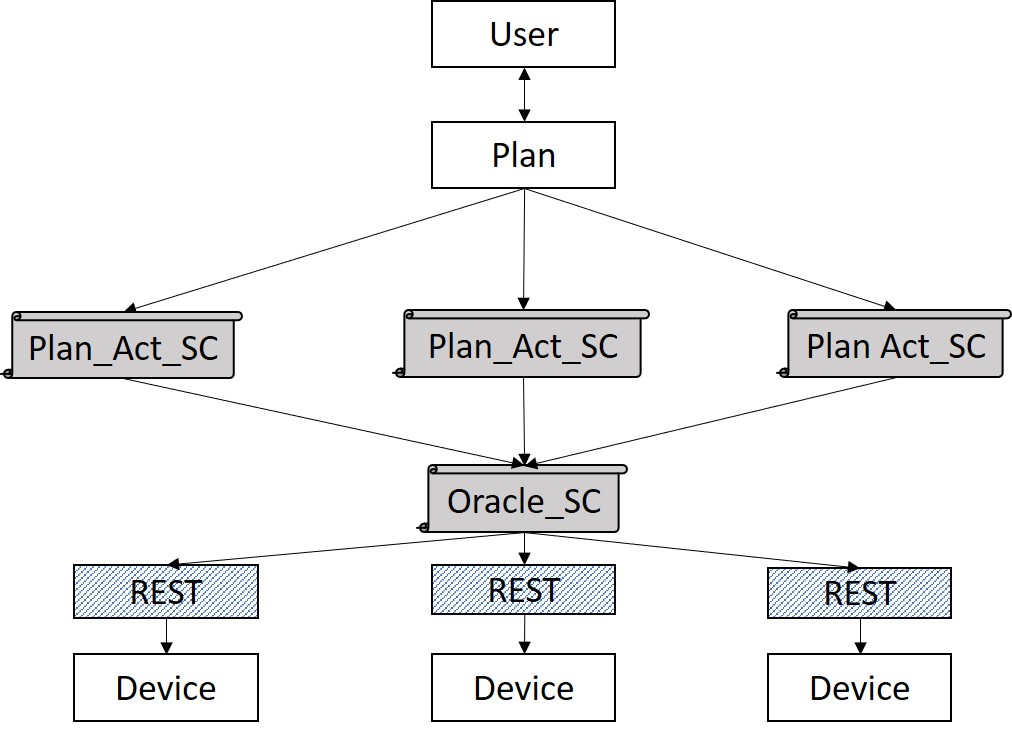}
	\caption{Distributed plan execution with Smart Contracts. \label{fig:dist-plan-exec-sm}}
\end{figure}

The main ideas for the smart contracts for decentralized mode operation is similar to the centralized case, except that the Scheduler $Plan\_SC$ and $Act\_SC$ are now combined with action dependencies maintained and updated in the single $Plan\_Act\_SC$.
The architecture for the distribute plan execution with smart contracts is given in Fig.~\ref{fig:dist-plan-exec-sm}. As before, we have a Register smart contract which provides a function to map actions to the corresponding agents. This mapping is used to call the update function of the $Plan\_Act\_SC$ instances. We also use the same Oracle smart contract to access the predicates and execute the actions.


In the centralized case the Act smart contracts were interacting with the Plan smart contract and there was no communication among the Act instances. On the other hand, in the distributed case, the $Plan\_Act$ smart contract interact peer-to-peer. Each $Plan\_Act$ smart contract maintains a $completedList$ which is used to find if any action can be enabled at all. When an action is enabled for execution, as before, precondition and effect are verified through the Oracle. When an action completes, the update function of the Act instances in $Out(a)$ are called so that their $completedList$s are updated.

\section{Related Work}
There is very little published research work in the intersection of AI and blockchain. But there is immense interest and a large number of ‘magazine’ style articles out there that talk about the potential. \cite{treleavelen:2016} talks about AI used as a cognitive technology for user interaction and how that enables a blockchain based solution. The online article \cite{Francesco:2017} argues at a high level several benefits of blockchains to AI and vice versa. In a number of papers from \cite{consortium:2017} AI is suggested to analyze the effectiveness of past contracts and suggest new terms and conditions. 

Blockchains are used to market and incentivize data, AI models and algorithms. Decisions from AI algorithms are proposed to be kept in the blockchain for provenance since blockchain ensures integrity of the data. Blockchain is used in federated learning to ensure accountability and prevent data poisoning~\cite{deepchain:weng:2018}. However, it seems the exploration of the intersection of AI and blockchain is still very much in its early days.

\cite{ calvaresi:2018} is a recent survey of the state of the art in the application of Blockchains to multi-agent systems. However in most cases, existing works do not address the issues in plan execution. For example, in \cite{ kapitonov:2017} the focus in on multi-agent coordination and control in the context of UAVs using blockchains.

Most work is focused on multi-agents and distributed vs centralized planning or the issues in coordinating planning activity in a distributed setup. There is very little on execution time coordination of plans among multiple agents. For example,  \cite{deWeerdt:2009,bonisoli:2013, dimopoulos:2006,nissim:2010} survey several multi-agent planning methods including distributed plans. 
\cite{guzman:2014} talks about how individual agents in a multi-agent setup may deal with failure situations. However, there is very little on how to enforce coordinated execution, especially in untrusted environments. In this paper, we show probably the most direct use of smart contracts in an AI use case to date.

\section{Conclusion}
Intelligent cyber-physical systems can be implemented by integrating planning capability in agent software, which transforms the system to a multi-agent planning paradigm. However, the traditional methods for planning and plan execution in multi-agent system can fail when the agents are untrusted. In this paper, we propose a smart-contract based extension of the agents which ensure proper plan execution even in untrusted environments.

We have described two specific architectures - one for a centralized and the other for a decentralized mode of execution of distribution plans. The architectures depend only upon the underlying devices and their agents and are independent of any applications built on top of the CPS. Moreover, the proposed architectures suggest that smart contracts can be automatically generated from the derived plans, as a result of which the entire system of smart, intelligent agents can be fully automated.

Another point to be noted is the clean separation of the physical layer operations (sensing and actuation) from the contract logic through the oracle smart contracts. Due to this characteristic, it is envisaged that functions such as plan synthesis and replanning 
can be easily integrated in the system by accessing these external services through the Oracle.
\balance
\bibliographystyle{IEEEtran}
\bibliography{paper.bib}

\clearpage
\section{Appendix}

\subsection{Register Smart Contract}
\begin{lstlisting}[language=Solidity, caption={\emph{Register Action} API},label={lst:registersc},captionpos=b,,numbers=none]
pragma solidity 0.4.0;
contract Register_SC {
    event logDeployedContract(address tenant,string s);
 	mapping(uint256 => Act_SC) action_scAddrMap;
    
    function setAct(uint[] memory action_names) public {
        Act_SC addr=new Act_SC(action_names); // returns address of new contract
        for(uint256 x = 0; x < action_names.length; x++) {
            action_scAddrMap[action_names[x]]=addr;
        }
        logDeployedContract(addr,"new_contract_address");
    }
    
    function  getAct(uint256 action_name) public returns(Act_SC act_scAddr) {
        return action_scAddrMap[action_name];
    }
}
\end{lstlisting}

\subsection{Plan Smart Contract}
\begin{lstlisting}[language=Solidity, caption={\emph{Plan} API},label={lst:plansc},captionpos=b,numbers=none]
pragma solidity ^0.4.0;
import "Register.sol";
contract Plan_SC{
    event loglist(uint256[] x,string s);    
    event logexist(uint256[] x,uint s,bool b);
    
    using ListInteger for *;
    uint  action_count;
    uint256[][6]  dag;
    uint256  CompletedListLength;
    function Plan_SC(){//DAG Creation
        action_count=6;
        dag[0]=[1]; dag[1]=[2];
    	dag[2]=[3,1]; dag[3]=[4,2];    
		dag[4]=[5,2,3]; dag[5]=[6,5];            
        }
    function runPlan(Register_SC register_addr){
        while(CompletedListLength!=action_count)
        for(uint256 x = 0; x < dag.length; x++) {
            uint current;bool _success;                 
            if(dag[x].getSize()==1){
            current =dag[x][0]; 
            _success=dispatchNext(register_addr,current);
            }
            if(_success){
                for(uint256 y = 0; y < dag.length; y++) {            
                    if(exist(dag[y],current))
			dag[y].removeByValue(current);                    			}
            }
        }
    }
    function dispatchNext(Register_SC register_addr, uint256 Action_id) returns(bool _success){
        Act_SC act_addr=register_addr.getAct(Action_id);
        bool success=act_addr.execute(Action_id);
        uint256[] memory res=act_addr.getCompletedList();
        CompletedListLength=res.length;
        loglist(res,"completed_list_updated");
        return success;
    }    
}
\end{lstlisting}

\subsection{Act Smart Contract}
\begin{lstlisting}[language=Solidity, caption={\emph{Act} API},label={lst:actsc},captionpos=b,numbers=none]
pragma solidity ^0.4.0;
import "github.com/oraclize/ethereum-api/oraclizeAPI.sol";
contract Act_SC  is usingOraclize{
    event display(string x);
    event display(string x,uint[] s);
    event abort(string s,uint256 actionID);
    event completed(string s,uint256 actionID);
    event displayActionName(uint256 x);
		struct Action{
        uint256  Action_id;
        string Action_URI;
        bytes32 [] precond;
        bytes32[] effects;
    }    
    Action private action_1,....,action_6;
    mapping(uint256 => Action) actionid_ActionMap;
    uint256[]  public completedList;
    
    function Act_SC(uint[] memory action_names){
        for(uint256 x = 0; x < action_names.length; x++) 	{
        if(action_names[x]==1){
        action_1.Action_id=action_names[x];
        action_1.Action_URI="http://a1.com";
        action_1.precond.push("precond1_0");        
        action_1.effects.push("effects1_0");
        actionid_ActionMap[1]=action_1;
        }//Similarly initialise other actions
    }        
}
    function getCompletedList() external returns(uint256[] list) {
        return completedList;
    }
    function getPreCondition(uint256 Action_id) public returns (bytes32[] preCond){
        return actionid_ActionMap[Action_id].precond;
    }
    function getEffect(uint256 Action_id) public returns(bytes32[] effect) {
          return actionid_ActionMap[Action_id].effects;
    }
    function checkPreCondition(bytes32[] preCond,bytes32[] preCond_orac) returns(bool cond) {
        return true;
    }
    function checkEffect(bytes32[] eff,bytes32[] eff_orac) public returns(bool cond) {
        return true;
    }
    //oraclize start from here...    
    bytes32 [] orac_res;// for storing preCond and effects
    function __callback(bytes32 myid, bytes32 [] result) {
        if (msg.sender != oraclize_cbAddress()) throw;
        orac_res = result;
    }
    function getPredicatesValues(uint256 Action_id,string s)returns(bytes32[] effect){
       string action_url=actionid_ActionMap[Action_id].Action_URI;
       oraclize_query("URL", action_url);// will call __callback
       return orac_res;
    }
    function execute(uint256 Action_id) external returns(bool success){
        bytes32[] memory preCond=getPreCondition(Action_id);
        bytes32[] memory preCond_orac=getPredicatesValues(Action_id,"preCond"); 
        if(checkPreCondition(preCond,preCond_orac)){
            display("callAPI(action_api_map[actionID])");
        }
        else{
            abort("Transaction_Rollback",Action_id);
            return false;
        }
        bytes32[] memory eff=getEffect(Action_id);
        bytes32[] memory eff_orac=getPredicatesValues(Action_id,"eff"); 
        if(checkEffect(eff,eff_orac)){
            completedList.push(Action_id);
            completed("Action_Completed", Action_id);
            return true;
        }
        else
            abort("Transaction_Rollback",Action_id);
       return false;
    }     
}
\end{lstlisting}

\end{document}

%% file: solidity-highlighting.tex
\usepackage{listings, xcolor}

\definecolor{verylightgray}{rgb}{.97,.97,.97}

\lstdefinelanguage{Solidity}{
	keywords=[1]{anonymous, assembly, assert, balance, break, call, callcode, case, catch, class, constant, continue, contract, debugger, default, delegatecall, delete, do, else, emit, event, export, external, false, finally, for, function, gas, if, implements, import, in, indexed, instanceof, interface, internal, is, length, library, log0, log1, log2, log3, log4, memory, modifier, new, payable, pragma, private, protected, public, pure, push, require, return, returns, revert, selfdestruct, send, storage, struct, suicide, super, switch, then, this, throw, transfer, true, try, typeof, using, value, view, while, with, addmod, ecrecover, keccak256, mulmod, ripemd160, sha256, sha3}, 
	keywordstyle=[1]\color{blue}\bfseries,
	keywords=[2]{address, bool, byte, bytes, bytes1, bytes2, bytes3, bytes4, bytes5, bytes6, bytes7, bytes8, bytes9, bytes10, bytes11, bytes12, bytes13, bytes14, bytes15, bytes16, bytes17, bytes18, bytes19, bytes20, bytes21, bytes22, bytes23, bytes24, bytes25, bytes26, bytes27, bytes28, bytes29, bytes30, bytes31, bytes32, enum, int, int8, int16, int24, int32, int40, int48, int56, int64, int72, int80, int88, int96, int104, int112, int120, int128, int136, int144, int152, int160, int168, int176, int184, int192, int200, int208, int216, int224, int232, int240, int248, int256, mapping, string, uint, uint8, uint16, uint24, uint32, uint40, uint48, uint56, uint64, uint72, uint80, uint88, uint96, uint104, uint112, uint120, uint128, uint136, uint144, uint152, uint160, uint168, uint176, uint184, uint192, uint200, uint208, uint216, uint224, uint232, uint240, uint248, uint256, var, void, ether, finney, szabo, wei, days, hours, minutes, seconds, weeks, years},	
	keywordstyle=[2]\color{teal}\bfseries,
	keywords=[3]{block, blockhash, coinbase, difficulty, gaslimit, number, timestamp, msg, data, gas, sender, sig, value, now, tx, gasprice, origin},	
	keywordstyle=[3]\color{violet}\bfseries,
	identifierstyle=\color{black},
	sensitive=false,
	comment=[l]{//},
	morecomment=[s]{/*}{*/},
	commentstyle=\color{gray}\ttfamily,
	stringstyle=\color{red}\ttfamily,
	morestring=[b]',
	morestring=[b]"
}